\documentclass[
reprint,           
superscriptaddress,
amsmath,           
amssymb,           
prd,               
notitlepage,       
longbibliography,  
floatfix,          
]{revtex4-1}

\usepackage{tensor}     
\usepackage{graphicx}   
\usepackage[
colorlinks=true,        
citecolor=blue,         
linkcolor=blue,         
urlcolor=blue           
]{hyperref}             
\usepackage{bm}         
\usepackage{xcolor}     

\newcommand{\newc}{\newcommand*}
\newc{\figurewidth}{3.2in}
\newc{\xbar}{\bar{x}}
\newc{\rhoeq}{\rho_{\rm{eq}}}
\newc{\zeq}{z_{\rm{eq}}}
\newc{\la}{\lambda}
\newc{\tla}{\tilde{\la}}
\newc{\dt}{\delta}
\newc{\Dt}{\Delta}
\newc{\vj}{\vec{j}}
\newc{\vl}{\vec{l}}
\newc{\hx}{\hat{x}}
\newc{\hy}{\hat{y}}
\newc{\bj}{\bm{j}}
\newc{\mJ}{\mathcal{J}}
\newc{\mP}{\mathcal{P}}
\newc{\ga}{\gamma}
\newc{\Msun}{M_\odot}
\newc{\app}{\approx}
\newc{\av}[1]{\langle #1 \rangle}
\newc{\eq}[1]{Eq.~\eqref{#1}}
\newc{\al}{\alpha}
\newc{\Xstar}{X_{\ast}}
\newc{\seq}{\sigma_{\rm{eq}}}
\newc{\fpbh}{f_{\rm{pbh}}}

\def\p{\partial}

\def\({\left(}
\def\){\right)}
\def\[{\left[}
\def\]{\right]}

\def\e{\begin{equation}}
\def\q{\end{equation}}
\def\m{\begin{eqnarray}}
\def\n{\end{eqnarray}}

\begin{document}

\title{
Effects of the surrounding primordial black holes on the merger rate of primordial black hole binaries}
\author{Lang Liu}
\email{liulang@itp.ac.cn} 
\affiliation{CAS Key Laboratory of Theoretical Physics,
Institute of Theoretical Physics, Chinese Academy of Sciences,
Beijing 100190, China}
\affiliation{School of Physical Sciences,
University of Chinese Academy of Sciences,
No. 19A Yuquan Road, Beijing 100049, China}
\author{Zong-Kuan Guo}
\email{guozk@itp.ac.cn}
\affiliation{CAS Key Laboratory of Theoretical Physics,
Institute of Theoretical Physics, Chinese Academy of Sciences,
Beijing 100190, China}
\affiliation{School of Physical Sciences,
University of Chinese Academy of Sciences,
No. 19A Yuquan Road, Beijing 100049, China}
\author{Rong-Gen Cai}
\email{cairg@itp.ac.cn}
\affiliation{CAS Key Laboratory of Theoretical Physics,
Institute of Theoretical Physics, Chinese Academy of Sciences,
Beijing 100190, China}
\affiliation{School of Physical Sciences,
University of Chinese Academy of Sciences,
No. 19A Yuquan Road, Beijing 100049, China}
\date{\today}
\begin{abstract}
We develop an analytic formalism for computing the merger rate of primordial black hole binaries with a general mass function by taking into account the torques by the surrounding primordial black holes and linear density perturbations.
We find that $\alpha=-(m_i+m_j)^2\p^2 \ln {\cal R}(m_i,m_j)/\p m_i \p m_j=36/37$ is independent of the mass function.
Moreover, the ratio of the merger rate density of primordial black hole binaries by taking into account the torques
by the surrounding primordial black holes to by the nearest primordial black hole
is independent of the masses of binaries.

\end{abstract}

\pacs{???}

\maketitle

Primordial black holes (PBHs) produced in the early Universe due to the collapse of large energy density fluctuations,
as a promising candidate for dark matter (DM),
have recently attracted much attention~\cite{Hawking:1971ei,Carr:1974nx,Carr:1975qj,Gao:2018pvq,Cai:2018rqf,Chen:2018rzo,Khlopov:2008qy,Belotsky:2018wph,Saito:2008jc}.

On the other hand, the gravitational wave events observed by the LIGO detectors~\cite{Abbott:2016blz}
can be explained by the coalescence of PBH binaries~\cite{Kashlinsky:2016sdv,Bird:2016dcv,Clesse:2016vqa,Sasaki:2016jop,Nakamura:1997sm,Ali-Haimoud:2017rtz,Bird:2016dcv,Nishikawa:2017chy,Raidal:2017mfl,Kocsis:2017yty,Sasaki:2018dmp,Kocsis:2017yty,Raidal:2017mfl}. At present, we do not know how to discriminate the PBH scenario from other astrophysical scenarios.
In Ref.~\cite{Kocsis:2017yty} a new method is proposed to test the PBH scenario.
A quantity $\alpha$ can be constructed from the mass distribution of the merger rate density per unit cosmic time and comoving volume  ${\cal R}(m_i,m_j)$.
It is found that the quantity,
$\alpha=-(m_i+m_j)^2\p^2 \ln {\cal R}(m_i,m_j)/\p m_i \p m_j$,
is independent of the black hole mass function.
In the PBH scenario, $\alpha$ is closed to unity.
For black holes from dynamical formation in a dense stellar system, $\alpha \approx 4$~\cite{OLeary:2016ayz}.
Therefore, it provides us a possibility to discriminate PBHs from astrophysical black holes.
However, in Ref.~\cite{Chen:2018czv} it is  pointed out that if $P(m)/m$ is a constant,
$\alpha=36/37$ which is consistent with the result derived in Ref.~\cite{Kocsis:2017yty}
while for a general mass function, the value of $\alpha$ deviates from $36/37$.
In this letter, we re-analyze the merger rate distribution of PBH binaries with a general mass function
by taking into account the torques by the surrounding PBHs and linear density perturbations.
We find that $\alpha=36/37$ for a general mass function.

In this letter, we use units of $c = G=1$.
Whenever relevant, we adopt the values of cosmological parameters from the \emph{Planck} measurements~\cite{Ade:2015xua}.
The scale factor $s(t)$ is normalized to unity at the matter radiation equality.

The probability distribution function of PBHs, $P(m)$, is normalized to be
\m
\int_0^\infty P(m)dm=1.
\n
The abundance of PBHs in the mass interval $(m, m+dm)$ is given by
\m
f P(m)dm, 
\n
where
$f\equiv\Omega_{\mathrm{pbh}}/\Omega_{\mathrm{m}}$ is the total abundance of PBHs in non-relativistic matter. The fraction of PBHs in DM is related to $f$ by $f_{\rm{pbh}}\equiv \Omega_{\rm{pbh}}/\Omega_{\rm{dm}} \approx f/0.85$. The average number density of PBHs in mass interval $(m, m+dm)$ at equality is given by
\m
n\(m\)dm= \frac{fP(m)dm \rho_{\rm{eq}}}{m},
\n
where $\rho_{\rm{eq}}$ is the energy density of the Universe at the matter radiation equality, 
and the comoving total average number density of PBHs $n_{T}$ is given by
\m
n_T\equiv f\rho_{\rm{eq}}\int_0^\infty {P(m)\over m}dm.
\n
For simplicity, we define $m_{pbh}$ as
\m
 \label{mpbh}
\frac{1}{m_{\rm{pbh}}}=\int_0^\infty {P(m)\over m} dm,
\n
and $F(m)$ as
\m
 \label{F(m)}
F\(m\)\equiv\frac{n\(m\)}{n_{T}}=P\(m\) \frac{m_{\rm{pbh}}}{m},
\n
where $F\(m\)$ is the fraction of the average number density of PBHs with mass $m$ in the total average number density of PBHs. It is easily obtained
\m
\int_0^\infty F(m)dm=1,
\n
which can be rewritten as
\m
\sum_{l=1}^{L} F(m_l) dm_l=1,
\n
where $m_1$ represents the minimum mass of PBH and $m_L$ represents the maximum mass of PBH.

Let us consider the condition two nearest PBHs with masses $m_i$ and $m_j$ decouple from the expanding Universe, 
assuming negligible initial peculiar velocities here and throughout. Considering the different scaling with time of the two competing effects (their gravitational attraction versus the expansion of the Universe), 
the equation of motion for their proper separation $r$ in Newtonian approximation is given by
\m
\label{eom1}
  \ddot{r} - \( \dot{H} + H^2 \) r + \frac{m_b}{r^2} \frac{r}{|r|} = 0,
\n
where the dot denotes differentiation with respect to the proper time and $m_b=m_i+m_j$ is total mass of the PBH binary. By defining $\chi \equiv r/x$, we can rewrite Eq.~\eqref{eom1} as
\m
  \chi'' + \frac{s h' + h}{s^2 h} \(s \chi' - \chi \) + \frac{1}{\la}
    \frac{1}{\(sh\)^2} \frac{1}{\chi^2} \frac{\chi}{|\chi|} = 0,
    \label{chi}
\n
where $x$ is the comoving distance between these two nearest PBHs and $h(s)\equiv H(s)/\({8\pi\over 3}\rhoeq\)^{1/2}=\sqrt{s^{-3}+s^{-4}}$. Primes denote differentiation with respect to scale factor $s$ and the dimensionless parameter $\la$ is
\m
\label{la}
  \la = \frac{8 \pi \rhoeq x^3}{3 m_b},
\n
 The solution of Eq.~\eqref{chi} in \cite{Ali-Haimoud:2017rtz} implies the semi-major axis $a$ of the formed binary is given by
\m
  a \approx 0.1 \la x .
      \label{axis}
\n

Assuming that the distribution of PBHs is random distribution, the probability distribution of the separation $x$ between two nearest PBHs with masses $m_i$ and $m_j$ and without other PBHs in the volume of ${4\pi \over 3}x^3$ is given by
\m
 \label{distribution}
&d{P}&\(m_i,m_j,x\)=F\( m_i\)dm_i 4\pi x^2 dx  n\(m_j\)dm_j
\nonumber \\
&\times& e^{- {4\pi \over 3}x^3 n(m_j) dm} \prod _{ m\neq m_j }^{ }{  e^{- {4\pi \over 3}x^3 n({m}) dm}}
\nonumber \\
&=&F\( m_i\)dm_i 4\pi x^2dx n\(m_j\)dm_j e^{- \int {4\pi \over 3}x^3 n_{m} dm}
\nonumber \\
&=&F\( m_i\)dm_i F\( m_j\)dm_j 4\pi x^2  n_{T}dx
e^{- {4\pi \over 3}x^3 n_{T}}\,.
 \nonumber  \\
\n
The average distance ${\hat x}_{ij}$  between two nearest PBHs with masses $m_i$ and $m_j$ is
\m
{\hat x}_{ij}&=&\int 4\pi x^2  x n_{T}
e^{- {4\pi \over 3}x^3 n_{T}} dx
 \nonumber  \\
&=& \frac{\rm Gamma(1/3)}{6^{2/3} \pi^{1/3}} n_{T}^{-1/3}
 \nonumber  \\
&\approx& 0.554 n_{T}^{-1/3}.
\n
Following Ref.~\cite{Ali-Haimoud:2017rtz}, we denoted by ${\bar x}_{ij}$ characteristic comoving separation between nearest PBHs with mass $m_i$ and $m_j$,
\m
\label{xbar}
{\bar x}_{ij}=\({3\over 4\pi}\)^{1/3} n_{T}^{-1/3}\approx 0.620 n_{T}^{-1/3},
\n
which is independent of the PBH binary masses. Therefore, from now on, we omit the subscript `$_{ij}$' unless it is necessary. Given a comoving separation $x$, we define the dimensionless variable $X$ as
\m
\label{X}
X \equiv (x /\overline{x})^3.
\n

When the two PBHs come closer and closer, the surrounding PBHs, especially the nearest PBH, will exert torques on the PBH binaries. As the result, the two PBHs avoid a head-on collision with each other. The tidal force will provide a angular momentum to prevent this system from direct coalescence. Here, we introduce a dimensionless angular momentum $j$ defined by
\m
j \equiv  \sqrt{1-e^2},
\n
where $e$ is the eccentricity of the binary at the formation time. From \cite{Ali-Haimoud:2017rtz}, the angular momentum generated by a PBH with mass $m$ at a comoving separation $y\ge x$ is given by
\m
  \bm{j} \approx 3 \frac{m}{m_b} \frac{x^3}{y^3}
    \(\hat{\bm{x}} \cdot \hat{\bm{y}}\) \(\hat{\bm{x}} \times \hat{\bm{y}}\).
\n
where $\hat{\bm{x}}$ is the unit vector along $\bm{x}$ and where $\hat{\bm{y}}$ is the unit vector along $\bm{y}$ .

We consider $N_l$ PBHs with mass $m_l$ uniformly distributed within a volume $V = \frac{4 \pi}{3} R^3$ and take the limit $N, V \rightarrow \infty$ at constant density $n(m_l) dm_l= N_l/V$. Therefore, the reduced angular momentum $\bm{j}_l$ resulting from $N_l$ PBHs with mass $m_l$ is given by
\m
\bm{j}_l \approx 3 \sum_{p=1}^{N_l} \frac{m_l}{m_b}\frac{x^3}{y_{l,p}^3} (\hat{\bm{x}} \cdot \hat{\bm{y}}_{l,p}) (\hat{\bm{x}} \times \hat{\bm{y}}_{l,p}).
\n
where $y_{l,p}$ is the comoving distance from the binary to the $p$-th PBH with mass $m_l$. The total reduced angular momentum $\bm{j}$ resulting from the surrounding PBHs with masses from $m_1$ to $m_L$ is given by
\m
\bm{j} = \sum_{l=1}^{L} \bm{j}_l \approx 3 \sum_{l=1}^{L} \sum_{p=1}^{N_l} \frac{m_l}{m_b}\frac{x^3}{y_{l,p}^3} (\hat{\bm{x}} \cdot \hat{\bm{y}}_{l,p}) (\hat{\bm{x}} \times \hat{\bm{y}}_{l,p}).
 \label{eq:j-sum}
\n

Using Eq.~\eqref{eq:j-sum}, the two-dimensional probability distribution for $\bm{j}$ is given by
\m
\frac{dP}{d^2j} = \underset{V \rightarrow \infty}{\lim} \prod_{l=1}^{L} \prod_{p=1}^{N_l} \int_V \frac{d^3 y_{l,p}}{V}   \delta_{\rm D} \left[\bm{j} -  3 \sum_{n=1}^{L} \sum_{q=1}^{N_n} \frac{m_n}{m_b} \frac{x^3}{y_{n,q}^5} y_{n,q ||}~ \bm{y}_{n,q \bot}\right],
 \nonumber \\
\n
where $y_{||} \equiv \bm{\bm{y}} \cdot \hat{\bm{x}}$, $\bm{y}_{\bot} \equiv \hat{\bm{x}} \times \bm{y}$, and $\delta_{\rm D}$ is the two-dimensional Dirac function, which we rewrite as
\m
\delta_{\rm D}(\bm{X}) = \int_{\bm{k} \bot \hat{x}} \frac{d^2 k}{(2 \pi)^2} {\rm e}^{i \bm{k} \cdot \bm{X}}.
\n
We hence get
\m
\frac{dP}{d^2j} &=& \underset{V \rightarrow \infty}{\lim} \prod_{l=1}^{L} \prod_{p=1}^{N_l} \int_V  \frac{d^3 y_{l,p}}{V} \int \frac{d^2 k}{(2 \pi)^2} 
  \nonumber \\
&\times& {\rm e}^{i   \(\bm{k} \cdot \bm{j} -  3 \sum_{n=1}^{L} \sum_{q=1}^{N_n} \frac{m_n}{m_b} \frac{x^3}{y_{n,q}^5} y_{n,q ||}~ \bm{k} \cdot \bm{y}_{n,q \bot}\)}
  \nonumber \\
&=& \underset{V \rightarrow \infty}{\lim} \int \frac{d^2 k}{(2 \pi)^2} {\rm e}^{i \bm{k} \cdot \bm{j}}
  \nonumber \\
&\times& \prod_{l=1}^{L} \prod_{p=1}^{N_l} \int_V  \frac{d^3 y_{l,p}}{V} {\rm e}^{i   \(-  3 \sum_{n=1}^{L} \sum_{q=1}^{N_n} \frac{m_n}{m_b} \frac{x^3}{y_{n,q}^5} y_{n,q ||}~ \bm{k} \cdot \bm{y}_{n,q \bot}\)}
  \nonumber \\
&=&  \underset{V \rightarrow \infty}{\lim} \int \frac{d^2 k}{(2 \pi)^2} {\rm e}^{i \bm{k} \cdot \bm{j}} \[\int_V \frac{d^3 y}{V} {\rm e}^{i   \(-  3  \frac{m_1}{m_b} \frac{x^3}{y^5}y_{ ||}~ \bm{k} \cdot \bm{y}_{ \bot}\)}\]^{N_1}
  \nonumber \\
&\times& ...... \times \[\int_V \frac{d^3 y}{V} {\rm e}^{i   \(-  3  \frac{m_L}{m_b} \frac{x^3}{y^5}y_{ ||}~ \bm{k} \cdot \bm{y}_{ \bot}\)}\]^{N_L}
  \nonumber \\
&=&  \underset{V \rightarrow \infty}{\lim} \int \frac{d^2 k}{(2 \pi)^2}  {\rm e}^{i \bm{k} \cdot \bm{j}} \prod_{l=1}^{L}
\mathcal{I}_l^{N_l},
\n
where
\m
\mathcal{I}_l &\equiv& \int_V \frac{d^3 y}{V} \exp\left[- 3 i \frac{m_l}{m_b}  \frac{x^3}{y^5} y_{||} \bm{k} \cdot \bm{y}_{\bot}\right]\nonumber\\
 &=& 1 - \frac1{V} \int_V d^3 y \left\{1 - \exp\left[- 3 i  \frac{m_l}{m_b} \frac{x^3}{y^5} y_{||} \bm{k} \cdot \bm{y}_{\bot}\right]  \right\}.
  \nonumber \\
\n
When $V \rightarrow \infty$ the latter integral is convergent, then we arrive
\m
\label{Il}
&\underset{V \rightarrow \infty}{\lim}& \mathcal{I}_l^{N_l}
  \nonumber \\
 &=& \underset{V \rightarrow \infty}{\lim} \left\{ 1 - \frac1{V} \int d^3 y \left[1 - {\rm e}^{- 3 i  \frac{m_l}{m_b} \frac{x^3}{y^5} y_{||} \bm{k} \cdot \bm{y}_{\bot}} \right] \right\}^{n(m_l) dm_l V} \nonumber\\
&=& {\rm e}^{- n(m_l) dm_l \mathcal{J}_l},
\n
where
\m
\mathcal{J}_l &\equiv& \int d^3 y \left(1 - \exp\left[- 3 i  \frac{m_l}{m_b}\frac{x^3}{y^5} y_{||} \bm{k} \cdot \bm{y}_{\bot}\right] \right)
  \nonumber \\
  &=& \int d^3 y \left(1 - \exp\left[- 3 i  \frac{m_l}{m_b}\frac{k x^3}{y^3} (\hat{\bm{y}} \cdot \hat{\bm{x}}) (\hat{\bm{y}} \cdot \hat{\bm{k}})\right] \right).
    \nonumber \\
\n
By rescaling $y \rightarrow (1.5 k)^{1/3} x y$ and defining $v = 1/y^3$, the integral $\mathcal{J}_l$ becomes
\m
\label{l}
\mathcal{J}_l &=& 1.5 k x^3 \int d^3 y \left(1 - \exp\left[2i \frac{m_l}{m_b} \frac{1}{y^3} (\hat{\bm{y}} \cdot \hat{\bm{x}}) (\hat{\bm{y}} \cdot \hat{\bm{k}})\right] \right)
\nonumber\\
&=& 2 \pi k x^3 \int_0^{\infty} \frac{d v}{v^2} \int \frac{d^2 \hat{y}}{4 \pi} \left(1 - {\rm e}^{2i \frac{m_l}{m_b} v (\hat{\bm{y}} \cdot \hat{\bm{x}}) (\hat{\bm{y}} \cdot \hat{\bm{k}})} \right)
\nonumber\\
&=& 2 \pi k x^3 \int_0^{\infty} \frac{d v}{v^2} A_l(v),
\n
where
\m
\label{A(v)}
A_l(v)=\int \frac{d^2 \hat{y}}{4 \pi} \left(1 - {\rm e}^{2i \frac{m_l}{m_b} v (\hat{\bm{y}} \cdot \hat{\bm{x}}) (\hat{\bm{y}} \cdot \hat{\bm{k}})} \right).
\n
By using
\m
 \hat{\bm{y}} \cdot \hat{\bm{x}}&=&\sin{\theta} \cos{\phi} , \nonumber \\
 \hat{\bm{y}} \cdot \hat{\bm{k}}&=&\sin{\theta} \sin{\phi},
\n
we can get
\m
\label{yxyk}
 (\hat{\bm{y}} \cdot \hat{\bm{x}}) (\hat{\bm{y}} \cdot \hat{\bm{k}})=\frac{\sin{2\phi}}{2}(1-\mu^2),
\n
where $\mu=\cos{\theta}$. From \eqref{A(v)} and \eqref{yxyk} and using $d^2 \hat{y}=\sin{\theta} d\theta d\phi$, $A_l(v)$ is given by
\m
\label{A(v)2}
A_l(v) &=&  \int_0^{2 \pi} \frac{d \phi}{2\pi}\int_0^1 d \mu \left(1- \exp\left[\frac{i v m_l}{m_b} \sin(2 \phi) (1- \mu^2) \right]\right)\nonumber\\
&=& \int_0^1 d \mu \left(1 - J_0\left[\frac{v m_l}{m_b}(1-\mu^2)\right]\right),
\n
where $J_0$ is the zeroth-order Bessel function. Since $J_0(x) = 1+ \mathcal{O}(x^2)$ for $x \rightarrow 0$, we could compute the integral over $v$ first. From \eqref{l} and \eqref{A(v)2}, $\mathcal{J}_l$ is given by
\m
\mathcal{J}_l &=& 2 \pi k x^3 \int_0^1 d \mu \int_0^{\infty} \frac{dv}{v^2} \left(1 - J_0\left[\frac{v m_l}{m_b}(1-\mu^2)\right]\right) \nonumber\\
&=& 2 \pi k x^3  \frac{m_l}{m_b}\int_0^1 d \mu (1- \mu^2) \int_0^{\infty} \frac{du}{u^2} (1 - J_0(u))\,,
\n
where $u=\frac{v m_l}{m_b}(1-\mu^2)$. The last two integrals are analytic. By using
\m
\int_0^1 d \mu (1- \mu^2)=2/3,\int_0^{\infty} \frac{du}{u^2} (1 - J_0(u))=1,
\n
we can get a simple expression
\m
\mathcal{J}_l = \frac{4 \pi}{3} \frac{m_l}{m_b} x^3 k.
\n
So, \eqref{Il} becomes
\m
\underset{V \rightarrow \infty}{\lim} \mathcal{I}_l^{N_l}={\rm e}^{- n(m_l) \mathcal{J}_l}={\rm e}^{- n(m_l) dm_l \frac{4 \pi}{3} \frac{m_l}{m_b} x^3 k}.
\n
Since $m_l n(m_l)=\rho_l$ is the energy density of PBHs with mass $m_l$, we can get
\m
\sum_{l=1}^{L} \rho_l dm_l=\rho_{\rm{pbh}}=f \rhoeq.
\n 
So we can arrive
\m
\prod_{l=1}^{L} \mathcal{I}_l^{N_l}={\rm e}^{-\frac{4 \pi}{3} \frac{f\rhoeq}{m_b} x^3 k}={\rm e}^{-j_X k},
\n
where
\m
j_X=\frac{4 \pi}{3} \frac{f\rhoeq}{m_b} x^3=\frac{m_{\rm pbh}}{m_b} X .
\n
We hence arrive at the probability distribution
\m
\frac{dP}{dj} &=& 2 \pi j \frac{dP}{d^2 j} = j \int \frac{d^2 k}{2 \pi}  {\rm e}^{i \bm{k} \cdot \bm{j} - j_X k}
\nonumber\\
 &=&  j \int k dk J_0(k j) {\rm e}^{- j_X k}
 \nonumber\\
 &=& \frac{j j_X}{(j^2+{j_X}^2)^{3/2}}.
\n
For a given $X$,
\m
\label{dP_dj}
j \frac{dP}{dj}\Big{|}_X &=& \mathcal{P}(j/j_X), \ \ \ \ \mathcal{P}(\gamma) \equiv \frac{\gamma^2}{(1 + \gamma^2)^{3/2}}.
\n

Similar to \cite{Ali-Haimoud:2017rtz}, taking into account both the torques by the surrounding PBHs and density perturbations, 
we can rewrite the characteristic value of $j_X$ as
\m
j_X \approx \frac{m_{\rm pbh}}{m_b} \left(1 + \sigma_{\rm eq}^2 /f^2 \right)^{1/2} X,
\label{jX}
\n
where $\sigma_{\rm{eq}}\equiv \langle \delta_{\rm{eq}}^2 \rangle^{1/2}$ is the variance of density perturbations of the rest of DM 
at the matter radiation equality. Once the PBHs decouple from the expanding Universe and form a binary, they gradually shrink by gravitational radiation and finally merge. The coalescence time can be estimated as~\cite{Peters:1964zz}
\m   \label{coalescence time}
  t = \frac{3}{85} \frac{a^4}{ m_i m_j m_b} j^7.
\n
We can rewrite~\eqref{coalescence time} as
\m
\label{sim j}
j=(\frac{3}{85} \frac{a^4}{ m_i m_j m_b})^{-1/7} t^{1/7}.
\n
The differential probability distribution of $X$ and $t$ is given by
\m
\frac{d^2P}{dX dt} = \frac{dP}{dX} \frac{d P}{dt}\Big{|}_X = \frac{dP}{dX} \times \left[\frac{\partial j}{\partial t} \frac{d P}{dj}\Big{|}_X \right]_{j(t; X)}.
\n
From \eqref{distribution} and \eqref{X}, we can get
\m
dP/dX = {\rm e}^{-X} F\( m_i\)dm_i F\( m_j\)dm_j.
\n
Given that $j \propto t^{1/7}$, $\partial j/\partial t = j/(7 t)$. By using~\eqref{dP_dj}, we can get
\m
\frac{d^2P}{dX dt} = \frac1{7 t} {\rm e}^{-X} F\( m_i\)dm_i F\( m_j\)dm_j\mathcal{P}\left(\gamma_X\right)\,,
\n
where
\m
\gamma_X \equiv \frac{j (t;X)}{j_X}.
\n
From Bayes' theorem, we get the probability distribution of $X$ for binaries merging after a time $t$
\m
\frac{dP}{dX} \Big{|}_{t} \propto \frac{d^2P}{dX dt}\Big{|}_{t}  \propto {\rm e}^{-X} \mathcal{P}\left(\gamma_X\right)\,.
\n

We now find the value of $X_*$ to satisfy the probability is maximized. Since $X_* \ll 1$,  we approximate ${\rm e}^{-X} \approx 1$. The equation we need to solve is
\m
0 = \frac{\partial}{\partial X} \left[\frac{dP}{dX} \Big{|}_{t}\right]_{X_*} \propto \mathcal{P}'(\gamma_{X_*}) \frac{\partial \gamma_X}{\partial X}.
\n
Since $\gamma_X$ is strictly monotonic, this implies $\mathcal{P}'(\gamma_{X_*}) = 0$, then we get 
\m
\label{j*}
j(t; X_*) = \sqrt{2} j_{X_*}.
\n
By solving \eqref{jX}, \eqref{sim j} and \eqref{j*}, the most probable value of $X$ for binaries merging at time $t$ is given by
\m
X_* &\approx& 2.12 f^{16/37} m_i^{3/37} m_j^{3/37} (m_i+m_j)^{36/37} t^{3/37}
 \nonumber\\
&\times& m_{\rm pbh}^{-1} \rhoeq^{4/37}  (1 + \frac{\sigma_{\rm eq}^2}{f^2})^{-21/74}.
 \label{eq:Xstar}
\n
The probability distribution of the time of PBH merger with mass $m_i$ and $m_j$ is given by
\m
\frac{dP}{dt} &=& \int dX \frac{d^2P}{dX dt}= \frac1{7 t} F\( m_i\)dm_i
 \nonumber\\
 &\times& F\( m_j\)dm_j \int dX {\rm e}^{-X} \mathcal{P}(\gamma_X) .
 \nonumber\\
\n
Since the integrand peaks at $X_* \ll 1$, we get ${\rm e}^{-X} = 1$
By using $\gamma_X \propto X^{-37/21}$, and $\gamma_{X_*} = \sqrt{2}$, we can find
\m
\int dX \mathcal{P}(\gamma_X) &=& \frac{21}{37} \frac{X_*}{\sqrt{2}} \int d \gamma (\gamma/\sqrt{2})^{-58/37} \mathcal{P}(\gamma) \nonumber\\
&\approx& 0.59 ~ X_*.
\n
The total probability distribution of the time of merger is given by
\m
\frac{dP_{T}}{dt}=\int \int 0.084 \frac{X_*}{t} F(m_i) F(m_j) dm_idm_j.
\n
The merger rate per unit volume at time $t$ is obtained from
\m
R(t)=\frac{d N_{\rm merge}}{dt dV} =\frac{1}{2}\frac{n_{T}}{(1+\zeq)^{3}} \frac{dP_{T}}{dt}\,,
\n
where the factor $1/2$ account for that each merger event involves two PBHs and $z_{\rm eq}\simeq 3400$ is the redshift at the matter radiation equality. Finally, we arrive at
\m
R(t)=\int \int  \mathcal{R}(m_i,m_j,t) dm_idm_j,
\n
where
\m
\label{R-d}
\mathcal{R}(m_i,m_j,t)&=&F\( m_i\) F\( m_j\) f^{\frac{53}{37}} (1 + \frac{\sigma_{\rm eq}^2}{f^2})^{-21/74}
 \nonumber\\
 &\times& 1.94\times 10^{6} \times \({ \Msun}\)^{{32 \over 37}} \({t\over t_0}\)^{-{34\over 37}} \({ m_i m_j}\)^{{3 \over 37}}
  \nonumber\\
 &\times& (m_i+m_j)^{\frac{36}{37}} m_{\rm pbh}^{-2}\,,
\n
which can be interpreted as the merger rate density in unit of Gpc$^{-3}$ yr$^{-1}$$\Msun^{-2}$.
From \eqref{R-d}, we confirm the result derived in~\cite{Kocsis:2017yty}, without considering the merger history of PBHs, $\alpha=-(m_i+m_j)^2\p^2 \ln \mathcal{R}(t,m_i,m_j)/\p m_i \p m_j=36/37$, which is independent of the PBH mass function, by taking into account the torques by the surrounding PBHs.

Only accounting for tidal torquing by the nearest PBH, the merger rate per unit volume at time $t$ is given by\cite{Liu:2019rnx}
\m
\hat{R}(t)=\int \int  \mathcal{\hat{R}}(m_i,m_j,t) dm_idm_j,
\n
where
\m
&\mathcal{\hat{R}}&(m_i,m_j,t) = \int F\( m_l\) \({ m_l}\)^{-{21 \over 37}} dm_l
\nonumber \\
& \times& 9.46\times 10^{5} \times \({ \Msun}\)^{{32 \over 37}} \({t\over t_0}\)^{-{34\over 37}} \({ m_i m_j}\)^{{3 \over 37}}
\nonumber \\
 & \times& F\( m_i\) F\( m_j\)  \({ m_{\rm pbh}}\)^{-{53 \over 37}} \({ m_i+m_j}\)^{{36 \over 37}}
 \fpbh^{53 \over 37}\,,
\n
which can be interpreted as the merger rate density in unit of Gpc$^{-3}$ yr$^{-1}$$\Msun^{-2}$.
As a comparation, we define $\lambda$ as
\m
\lambda \equiv \frac{\mathcal{\hat{R}}(m_i,m_j,t)}{\mathcal{R}(m_i,m_j,t)}\,.
\n
By using $f_{\rm pbh}=f/0.85$, $\lambda$ is given by
\m
\lambda=0.61 \times \int F\( m_l\) \({ m_l}\)^{{21 \over 37}} dm_l \({ m_{\rm pbh}}\)^{-{21 \over 37}} (1 + \frac{\sigma_{\rm eq}^2}{f^2})^{21/74}\,,
\nonumber \\
\n
which is independent of $m_i,m_j$and $t$.

Now let us consider two typical PBH mass functions. 
One is the monochromatic case~\cite{Sasaki:2016jop,Bird:2016dcv,Nishikawa:2017chy},
\m
P(m)=\delta(m-M).
\n
The other takes the power-law form as follows~\cite{Carr:1975qj}
\m\label{power}
 P(m) = {\beta-1\over M} \({m\over M}\)^{-\beta}\,,
\n
with $m\geq M$ and $\beta>1$. In the monochromatic case, we arrive at
\m
\lambda=0.61 \times (1 + \frac{\sigma_{\rm eq}^2}{f^2})^{21/74}\,.
\n
Neglecting the density perturbation, $\lambda=0.61$ is independent of the mass of PBH. In the power-law case, $\lambda$ is given by
\m
\lambda=0.61 \times \frac{37 \beta^{58/37}}{(\beta-1)^{21/37}(21+37\beta)}(1 + \frac{\sigma_{\rm eq}^2}{f^2})^{21/74}\,.
\n
Neglecting the density perturbation, in the case $\beta>1.05$, $\lambda\sim {\cal O}(1)$ is almost independent on the mass function of PBH.

\begin{figure}[htbp!]
\centering
\includegraphics[width = 0.48\textwidth]{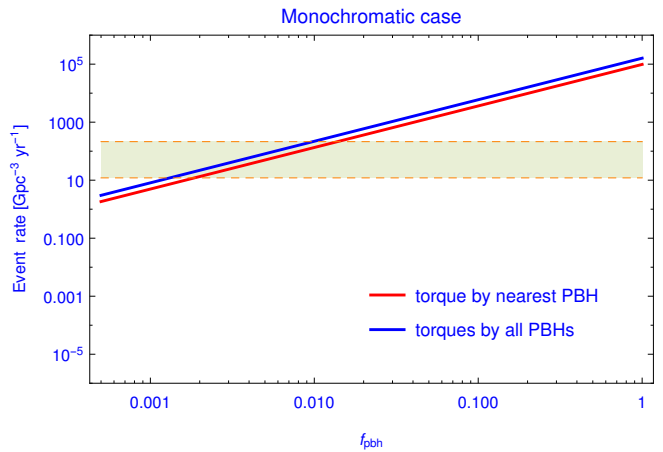}
\caption{\label{fig:Monochromatic}
Event rate of mergers of monochromatic PBH binaries whose mass is $30 \Msun$ as a function of the dark matter in PBHs. The blue line is the case for taking into  account the torques by all PBHs and the red  line is the case for taking into  account the torque by the nearest PBH. The event rate as $R=12\sim 213$ Gpc$^{-3}$ yr$^{-1}$ estimated by the LIGO-Virgo Collaboration is shown as the shaded region colored orange~\cite{Abbott:2017vtc}. }
\end{figure}

\begin{figure}[htbp!]
\centering
\includegraphics[width = 0.48\textwidth]{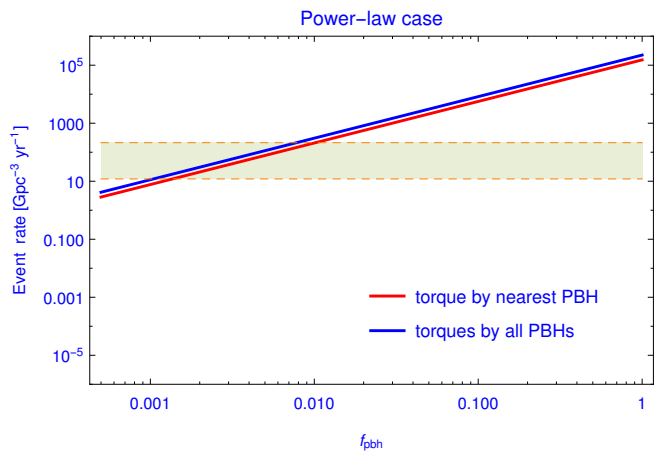}
\caption{\label{fig:Power}
By taking $\beta=2.3$, $M=5\Msun$, the merger rate of PBH binaries at present with $m_1,\ m_2\geq 5M_\odot$ and $m_1+m_2\leq 100M_\odot$. The blue line is the case for taking into  account the torques by all PBHs and the red  line is the case for taking into account the torque by the nearest PBH. The event rate as $R=12\sim 213$ Gpc$^{-3}$ yr$^{-1}$ estimated by the LIGO-Virgo Collaboration is shown as the shaded region colored orange~\cite{Abbott:2017vtc}.  }
\end{figure}

Fig.~\ref{fig:Monochromatic} and Fig.~\ref{fig:Power} show that the merger rate constrained by LIGO/VIRGO could be explained by mergers of PBH binaries. Here we take $M=30\Msun$ for the monochromatic case and $M=5\Msun,\beta=2.3$ for the power-law case. In the monochromatic case, LIGO/VIRGO implies that $0.0019 \lesssim f_{\rm{pbh}}\lesssim 0.014$ for taking into account the torque by the nearest PBH and $0.0013 \lesssim f_{\rm{pbh}}\lesssim 0.0098$ for taking into account the torques by all PBHs. In the power-law case, LIGO/VIRGO implies that $0.0014 \lesssim f_{\rm{pbh}}\lesssim 0.010$ for taking into  account the torque by the nearest PBH and $0.0010 \lesssim f_{\rm{pbh}}\lesssim 0.0077$ for taking into account the torques by all PBHs. Such an abundance of PBHs satisfies current constraints from other observations \cite{Chen:2016pud,Green:2016xgy,Schutz:2016khr,Wang:2016ana,Gaggero:2016dpq,Ali-Haimoud:2016mbv,Blum:2016cjs,Horowitz:2016lib,Kuhnel:2017pwq,Inoue:2017csr,Carr:2017jsz,Green:2017qoa,Guo:2017njn,Poulin:2017bwe}.

So far, several gravitational wave events from black hole binary mergers have been detected by LIGO/VIRGO collaboration, such as GW150914~\cite{Abbott:2016blz}, GW151226~\cite{Abbott:2016nmj}, GW170104~\cite{Abbott:2017vtc}, GW170608~\cite{Abbott:2017gyy} and
GW170814~\cite{Abbott:2017oio}. 
One of the most important question is how to discriminate PBHs and astrophysical black holes. 
In this letter, we develop a formalism for calculating the merger rate distribution of PBH binaries with a general mass function by taking into account the torques by the surrounding PBHs and linear density perturbations. 
We find $\alpha=36/37$ is independent of the mass function of PBHs.  
The ratio of the merger rate density of PBHs by taking into account the torques by the surrounding PBHs to the nearest PBH is independent of $m_i$ and $m_j$. 
We apply our formalism to two specific examples, the monochromatic and power-law cases. 
In these cases, three body approximation is a good approximation.
In the future, more and more coalescence events of black hole binaries will be detected by LIGO/VIRGO.
This will provide more information of the merger rate distribution of black hole binaries to test the PBH scenario.

\centerline{\bf Note added}
In finishing this letter, we find a parallel independent work \cite{Raidal:2018bbj} that has some overlap with our calculation of the merger rate of primordial black hole binaries.

\begin{acknowledgments}
We thank Jing Liu, Yu-Tian Shen, Zu-Cheng Chen, Qing-Guo Huang for helpful discussions. This work is supported in part by the National Natural Science Foundation of China Grants
No.11575272, No.11690021, No.11690022, No.11851302, No.11375247 and No.11435006,
in part by the Strategic Priority Research Program of the Chinese Academy of Sciences Grant No. XDB23030100,
No. XDA15020701 and by Key Research Program of Frontier Sciences, CAS.\end{acknowledgments}


\bibliography{Rpbh_v2}

\end{document}